\documentclass[runningheads]{llncs}

\usepackage{epic,eepic,epsf}
\usepackage{subfigure}
\usepackage{wrapfig}
\usepackage{caption2}
\usepackage{algorithmic}
\usepackage{algorithm}
\usepackage{subfigure}

\usepackage{stmaryrd}

\newcommand{\ol}{\setlength{\itemsep}{0pt.}\begin{enumerate}}
\newcommand{\eol}{\end{enumerate}\setlength{\itemsep}{-\parsep}}

\newcommand{\comment}[1]{}

\newcommand{\remove}[1]{}

\renewcommand{\qed}{\hfill\rule{2mm}{2mm}}

\begin{document}
\mainmatter              % start of the contributions
\title{RFID Authentication\\{\Large Efficient Proactive Information Security within Computational Security}\thanks{Partially supported by Microsoft, IBM, NSF, Intel, Deutsche Telekom, Rita Altura Trust Chair in Computer Sciences, Intel, vaatat and Lynne and William Frankel Center for Computer Sciences.}\\ 
\vspace*{0.4cm}
%\small{{\rm(CS BGU Technical Report \#08, July 17th 2007)}}
\vspace*{-0.4cm}
 }

\titlerunning{RFID Authentication}  % abbreviated title (for running head)
%                                     also used for the TOC unless
%                                     \toctitle is used
%
\author{Shlomi Dolev\inst{1} \and Marina Kopeetsky\inst{2} \and Adi Shamir\inst{3}}
\authorrunning{Shlomi Dolev, Marina Kopeetsky and Adi Shamir}   % abbreviated author list (for running head)
%
%%%% modified list of authors for the TOC (add the affiliations)
\tocauthor{Shlomi Dolev (Ben-Gurion University),
Marina Kopeetsky (Sami Shamoon College of Engineering),
Adi Shamir (Weizmann Institute)}
\institute{
Department of Computer Science, Ben-Gurion University of the
Negev, Beer-Sheva, 84105, Israel \email{dolev@cs.bgu.ac.il}
%\thanks{Partially supported by Microsoft, IBM, NSF, Intel,
%        Deutsche Telekom, Rita Altura Trust Chair in Computer Sciences,
%        Intel, vaatat and Lynn and William Frankel Center for Computer Sciences.}
\and
Department of Software Engineering, Sami-Shamoon College of Engineering, Beer-Sheva, 84100,
 Israel \email{marinako@sce.ac.il}
\and 
Department of Computer Science \& Applied Mathematics, Weizmann Institute of Science, Rehovot 76100, Israel \email{adi.shamir@weizmann.ac.il}
}

\maketitle              % typeset the title of the contribution

%%%%%%%%%%%%%%%%%%%%%%%%%%%%%%%%%%%%%%%%%%
%\renewcommand{\thefootnote}{\fnsymbol{footnote}}

\begin{abstract}
We consider repeated communication sessions between a RFID Tag 
(e.g., Radio Frequency Identification, RFID Tag) and a RFID Verifier.
A proactive information theoretic security scheme is proposed.
The scheme is based on the assumption
that the information exchanged during at least one 
of every $n$ successive communication sessions is not exposed 
to an adversary. The Tag and the Verifier maintain
a vector of $n$ entries that is repeatedly refreshed by 
pairwise $xoring$ entries, with a new 
vector of $n$ entries that is randomly chosen by the Tag
and sent to the Verifier as a part of each communication session. 

The general case in which the adversary does not listen
in $k \geq 1$ sessions among any $n$ successive 
communication sessions is also considered.
A lower bound of $n \cdot (k+1)$ for the number of 
\remove{new} random numbers used during any $n$ successive communication 
sessions is proven. In other words, we prove that an algorithm must use at least $n \cdot (k+1)$ new random numbers during any $n$ successive communication sessions. Then a randomized scheme that
uses only $O(n \log n)$ new random numbers is presented.

A computational secure scheme which is based on the information theoretic secure 
scheme is used to ensure that even in the case that the adversary 
listens in all the information exchanges, the communication 
between the Tag and the Verifier is secure.\\
%In particular, the scheme can be used in the domain of remote controls (e.g., for cars). 
\end{abstract}

\renewcommand{\thefootnote}{\arabic{footnote}}

\section{Introduction}
\label{sec:Introduction}
RFID tag is a small microchip, supplemented with an antenna, that transmits a unique identifier in response to a query by a reading device. The RFID technology is designed for the unique identification of different kinds of objects. According to \cite{SWE03}, RFID communication systems are composed of three major elements: \\
(a) RFID Tag carries object identifying data; (b) RFID Verifier interfaces with Tags to read or write Tag data; (c) the back-end database aggregates and utilizes Tag data collected by Verifiers.
 
The RFID Verifier broadcasts an RF signal to access data stored on the tags that usually includes a unique identification number.
RFID tags are designed as low cost devices that use cheap radio transmission media. Such tags have no (or a very limited) internal source of power. However they receive their power from the reading devices. 
The range of the basic tags transmission is up to several meters. 
Possible applications of the RFID devices include: RFID-enabled banknotes, libraries, passports, pharmaceutical distribution of drugs, and organization of the automobile security system or any key-less entry system.  Nevertheless, the wide deployment of RFID tags may cause new security and privacy protecting issues. RFID tags usually operate in insecure environment. The RFID tag privacy may be compromised by the adversary that extracts unencrypted data from the unprotected tags. RFID tags are limited devices that cannot support complicated cryptographic functions. Hence, there is growing interest in achieving high security and privacy level for the RFID devices, without usage of computationally expensive encryption techniques. 

The focus of our paper is the authentication schemes for passive RFID tags. 
We present new proactive and cost effective information theoretic and 
computationally secure authentication protocols for RFIDs.
The main scope is one sided authentication, where the Verifier
has to identify the Tag. Such (non mutual) one sided 
authentication is useful in applications in which the 
Tag may have other means to identify (that it is communicating
with) the desired Verifier (say by being geographically close
to the Verifier). 
Note that a simple extension in which symmetric authentication 
scheme obtains mutual authentication is also presented in the sequel.
The protocol that copes with the Intruder-in the Middle-Attack (IIMA) is introduced as the extended version of the computationally secure protocol. 

%We also exclude the possibility of 
%man-in the-middle attacks, having similar applications
%in minds, where the Tag may identify the existence
%of a man-in the-middle. Still, we conclude and suggest ways 
%to cope with these limitations.
 
\noindent{\bf Background and related work.}\\
$\bullet$ \noindent{\bf Security protocols based on standard cryptographic techniques.}\\
A brief introduction to RFID technology appears in
\cite{SWE03} where potential security and privacy risks are described. Schemes for providing desired security properties in the unique setting of low-cost RFID devices are discussed in \cite{SWE03}. 
\remove{The authors of \cite{SWE03} depict several advantages of the RFID tags over traditional optical bar codes. Unlike the optical bar codes, RFID Verifiers are able to read data automatically through non-conducting material at a rate of several hundred tags per second and from a distance of several meters. The authors state that low-cost smart RFID tags may become an efficient replacement for optical bar codes.}
The main security risks stated in that paper are the violations of ``location privacy'' and denial of service that disables the tags. With the RFID resource constraints in mind, the cryptography techniques proposed in \cite{SWE03} for developing the RFID security mechanisms are:	
(a) a simple access mechanism based on hardware-efficient one-way hash functions, low-cost traditional symmetric encryption schemes, randomizing tag responses based on random number generator;
(b) integrating RFID systems with a key management infrastructure. Regardless of the mechanisms used for privacy and access control, management of tag keys is an important issue. The new challenge in the RFID system design is to provide access control and key management tools compatible with the tags cost constraints.

An adversary model adapted to RFID protocols is introduced in \cite{Avoine}. Many existing privacy protecting RFID protocols are examined for their \textit{traceability}. \textit{Traceability} is defined as the capability of the adversary to recognize a tag which the adversary has already seen, at another time or in another location \cite{Avoine}.  
The traceability is stated as a serious problem related to the privacy protection in the RFID systems. The paper concludes that in a realistic model, many protocols are not resistant to traceability.\\
$\bullet$ \noindent{\bf Security protocols based on low cost arithmetic computations.}\\
The research survey in \cite{J205} examines different approaches proposed by researches for providing privacy protection and integrity assurance in RFID systems. In order to define the notions of ``secure'' and ``private'' for RFID tags, a formal model that characterizes the capabilities of potential adversaries is proposed. The author states that it is important to adapt RFID security models to cope with the weakness of the RFID devices. A few weak security models which reflect real threats and tag capabilities are discussed. A ``minimalist'' security model which serves low-cost tags is introduced in \cite{JM05}. The basic model assumption is that the potential RFID adversary is necessarily weaker than the one in traditional cryptography. Besides, such an adversary comes into scanning range of a tag only periodically. The minimalist model aims to take into account the RFID adversary characteristics. Therefore, this model is not perfect, but it eliminates some of the standard cryptographic assumptions that may be not appropriate for deployment in other security systems which are based on a more powerful adversary model. The author of \cite{JM05} states that standard cryptographic functionality is not needed to achieve necessary security in RFID tags. \\
\remove{
The paper \cite{LiCho07} proposes the low cost mutual authentication protocol. This protocol is based only on simple $xor$  computation and the issue of Partial Identification Number, which is in essence a sub string of Tag secret Identification Number.The proposed protocol supports major security features required for the RFID systems, such as mutual authentication, traffic encryption and privacy protection including the traceability problem.

The paper \cite{LWSP07} proposes a light-weight security protocol for RFID system. The designed protocol is based on one-time pad scheme and it satisfies the secrecy, data authenticity and data anonymity. The hardware efficient implementation is also presented.

A real lightweight mutual authentication protocol for low-cost RFID tags that satisfies adequate security level, is proposed in \cite{LMAP}. LMAP protocol is based on the use of pseudonyms, precisely in index-pseudonyms that determine the index of a table (a row) where all secret information about a tag is stored. LMAP's is implemented by means of low cost arithmetic operations such as bitwise $xor$, bitwise $or$, bitwise $and$, and addition $mod~2^{m}$.  
} 
$\bullet$ \noindent{\bf Protocols overview.}\\
Existing techniques and secure protocols proposed for implementation in existing RFID systems are described next.

An inexpensive RFID tag known as Electronic Product Code (EPC) tag is proposed in \cite{J105} to protect against RFID tag cloning. Basic EPC tags do possess features geared toward privacy protection and access control mechanisms, nevertheless they do not possess explicit authentication functionality. That is, EPC standards prescribe no mechanism for RFID-EPC verifiers to authenticate the validity of the tags they scan. The authors show how to construct tag-to-verifier and verifier-to-tag authentication protocols. 
 
However, the security analysis of the Digital Signature Transponder (DST) RFID tags is described in \cite{BGSJRS05}. The authors present in detail the successful strategy for defeating the security of an RFID device known as Digital Signature Transponder. The main conclusion of \cite{BGSJRS05} is that the DST tags are no longer secure due to the tags weakness caused by the inadequate short key length of 40 bits. Note that it is possible to increase the computational security level by increasing the length of the key. Still the resulting scheme will not be information theoretic secure but only computationally secure. Hence, it is of interest to design a proactive information theoretic secure scheme within computational secure scheme as we do in the sequel.

We also detail a new way to use watermarks technique to cope with IIMA
even for the case there is only one message sent during a communication
session (unlike \cite{JM05,DK07} where the exchange of three messages 
is required).
The new scheme is based on expanding each message to be a
codeword with error correcting
bits. Thus, enforcing an attacker to change at least a number
of bits equal to the minimal Hamming distance between 
two codewords. 
In addition watermarks bits are produced by pseudo-random sequence
and inserted in the message in specific locations defined by pseudo-random 
sequences. The operations for producing watermarked messages are
based only on $xor$ operations and the usage of
pseudo-random sequences, rather than using cryptographic hash functions.\\
Note that one can use symmetric 
authentication scheme to obtain mutual authentication of the
Tag and the Verifier. The mutual authentication scheme allows the Verifier to produce random bits for the Tag, as well. 
\remove{For example, we may
double the number of entries in the vectors of the Tag
and the Verifier and use the one entry to authenticate the
Tag and the next entry to authenticate the Verifier. }
Obviously, computational security ``envelop'' can be implemented
for the symmetric version as well, resulting in a proactive computational
secure symmetric scheme. We present in detail the one sided authentication, for readability purposes.

\noindent{\bf Our contribution.}
Our goal in this paper is to design new algorithms for providing 
authentication for the computationally limited basic RFID systems 
with a small amount of storage capability.

We propose a new security protecting model that is information theoretic and 
computationally secure. The security power of the basic and 
combined authentication protocols is provided by maintaining at 
the Tag and the Verifier's sides $n$-dimensional vector $ARV$. 
The appropriate vector-entry is used as the secret 
key for performing the authentication procedure by the RFID Tag. 
The vector $ARV$ is updated by performing $xor$ of corresponding entries with randomly chosen new $n$ dimensional vector at any communication session. Our work is mostly related to the schemes presented in \cite{DK} and filed as patent \cite{DKpatent}.
In this paper we present here an equivalent solution that maintains only
a vector of $n$ numbers instead of $O(n^{2})$ numbers. 
In addition we present a new algorithm that uses randomization
in order to reduce the communication during a session
from $O(n)$ numbers to $O(\log~n)$ numbers.
 
The basic information theoretic secure protocol $AP_{1}$ is based on the limited adversarial capabilities. The underlining assumption of this protocol is that the adversary is not listening in at least one of each $n$ successive interactions between the Tag and the Verifier. In essence, $AP_{1}$ protocol extends the ``minimalist'' security model in \cite{JM05} and the assumptions made in \cite{Bluetooth}. 
The underlying assumption of $AP_{1}$ is that each communication session 
is atomic. We mean that the adversary cannot modify part of the 
communication in a session.
The adversary may either listen in the communication during a session, or try to communicate (on behalf of the RFID Tag) during an entire session. This is a common situation in the case of personal identification, when an adversary cannot be present when the user is.
Compared with \cite{JM05} our scheme is not based on an oracle which provides the number of sessions in which the adversary listens in following the last refresh. Our scheme works when we do not know explicitly which session the adversary is not listening in. In other words, \cite{JM05} needs to identify such a session in order to renew the security level in this session while our scheme renews the security level without the need to identify such a particular secure session. Moreover, the security failure in a certain session does not bear on successful implementation of the next sessions since our algorithms are proactive.

A proactive information theoretic security scheme is proposed.
According to \cite{Amir} proactive security provides a method for maintaining the overall security of a system, even when individual components are repeatedly broken into and controlled by an attacker. The automated recovery of the security is provided by our scheme.
The scheme is based on the assumption
that the information exchanged during at least one 
of every $n$ successive communication sessions is not exposed 
to an adversary.
% The scheme is based on maintaining a vector of $n$ entries that is refreshed by 
The vector is refreshed by pairwise $xor-ing$ entries, with a new 
vector of $n$ entries that is randomly chosen by the Tag
and sent to the Verifier as a part of each communication session. 

The general case in which the adversary does not listen
in $k \geq 1$ sessions among any $n$ successive 
communication sessions is also considered.
We prove an $n \cdot (k+1)$ lower bound for the number of random numbers required when a deterministic version
of our scheme is used. The lower bound is on the number of 
new random numbers used during any $n$ successive communication 
sessions. In other words we prove that any deterministic algorithm will use at least $n \cdot (k+1)$ random numbers during any $n$ successive communication sessions. 
Then we present a randomized scheme that uses
only logarithmic in $n$ random numbers 
in each communication session, assuming the adversary
does not listen in a bounded fixed portion of 
any $n$ successive communication sessions.

The restriction imposed on the adversary is dropped in the combined proactive computational secure protocol $AP_{2}$ that operates successfully even if the adversary has gotten access to any number of successive interactions between the Tag and the Verifier. 
%$AP_{2}$ protocol does not follow the ``minimalist model'' proposed in \cite{JM05}. 
%There are only (reasonable) computation limitations on the 
%adversarial capabilities. 
%$AP_2$ does not rely on atomic sessions and it is resistant against active intruder-in the-middle attacks \cite{S02}.
\remove{In \cite{DK} there is an atomicity of session assumption. }
We extended $AP_{2}$ to a version $\widetilde{AP_{2}}$ that does not rely on atomic sessions and is computationally resistant against active IIMAs.
 The proactive combined computational secure protocol has several advantages.
 \\ \textbf{Low computational cost combined with a high security level.} Our algorithms continuously use random numbers generator as a source for preserving the security level (\cite{RNG}). Low computational power is required compared with the standard cryptographic techniques like stream and block ciphers. \\ 
\textbf{Protocols' robustness.} Our proactive computational secure protocol is not based on the refreshing procedure as suggested in \cite{JM05}. The refreshing procedure in \cite{JM05} provides the complete initialization of the protocol's secure parameters assuming there is an oracle that identifies sessions in which the adversary is not listening in. Namely, the refresh is done via a secure channel. 
%Moreover, the trusted party or RFID verifier in \cite{JM05} accesses the RFID system on a periodic basis refreshing the system. 
Our model provides high computational security level by involving a trusted party  only during initialization, without identifying a particular session as a secure session. \\
%\textbf{Security system reliability.} $AP_{1}$ does not rely on information concerning the specific session among consecutive $n$ sessions the adversary was listening in, as \cite{JM05} assumes in order to reinitialize the protocol.\\
\textbf{Functionality in the proactive mode}. 
\remove{According to \cite{Amir} proactive security provides a method for maintaining the overall security of a system, even when individual components are repeatedly broken into and controlled by an attacker. The automated recovery of the security protocol is provided in the proactive security model \cite{Amir}.}
   Any listening adversary's success and consequent protocol's security failure do not affect further functionality of the protocol. Recovery from a failure (assuming non fatal effect of failures) is automatic. That is to say, assuming that no fatal damage is caused when the adversary reveals the clear text, the future communication security is established. \\                                                                                                                        
\textbf{Possibility of proactive information theoretic security within computational security}. 
Our second protocol $AP_{2}$ assumes that if the adversary was not listening in at least one session among \textit{n} consecutive sessions between the RFID Tag and the RFID Verifier, the proposed protocol automatically becomes information theoretic and computationally secure and therefore the original security level is established. Thus, an adversary that starts processing the communication information in order to break the computational security based scheme, will have to start
from scratch after any session the adversary did not listen in. This fact can be used,
in turn, to reduce the number of random bits used with relation to an only
computational secure scheme. 
Assume that there is some probability for value of $n$ to be the correct number of sessions for which the adversary does not listen in at least one session. Assume further that there 
is a larger definite upper bound, $n' \geq n$, that may depend on 
a stricter consideration, (say battery lifetime). 
In such a case it is possible to tune the computational security level of $AP_{2}$ to fit 
the need to secure the sessions in which the 
protocol is not information theoretic secure, taking in account the probability that the adversary will indeed be present in $n'$ (or less) successive sessions.\\
\textbf{High level of the computational resistance against active IIMAs}. 
Security against IIMAs of the updated $\widetilde{AP_{2}}$ is 
achieved by means of the low cost $xor$-based techniques of the 
redundant coding \cite{rc} and digital watermarking \cite{DWM}. 
The techniques used by $\widetilde{AP_{2}}$ loosens the assumption on 
the atomicity of any session. 
A protocol that is resistant to IIMAs is proposed in \cite{JM05}. 
The protocol is based on the 
three-way mutual authentication procedure between the RFID Tag and the 
RFID Verifier. The protocol's computational security power is 
achieved by means of one-time pads that encapsulate the secret 
keys, and by the constant keys updating in each communication session. 
Another such protocol that is based on three message exchanges in each
session is proposed in \cite{DK07}. This protocol is provably 
secure based on the hardness of the Learning Parity in the Presence 
of Noise problem. Compared with \cite{JM05} and \cite{DK07} 
our $\widetilde{AP_{2}}$ protocol can be used for one way authentication
with only one message exchange for session, or two-way authentication using
two messages.
Thus, our scheme is also applicable in the cases in which
the RFID Verifier does not send messages to the RFID Tag. \\
%for example in the scope of automobile or any other key-less entry system. \\
%Moreover, the $\widetilde{AP_{2}}$ security parameters do 
%not determine and do not influence the RFID system reaction, 
%while in \cite{JM05} the number of queries that an adversary can 
%make in mounting IIMAs is bounded by the system parameters. \\
%\textbf{Extension to $k$ sessions that the adversary does not listens in} \\
We believe that our protocols are useful in several domains including remote keys, e.g., automobile security system, in particular the mutual authentication versions of our protocols. 
  
\noindent{\bf Paper organization.}
The formal system description appears in Section \ref{s:fsd}.
The basic information theoretic secure protocol $AP_{1}$ is introduced in Section \ref{s:isp}. 
The case in which the adversary does not listen in 
$k > 1$ sessions of any $n$ successive sessions is investigated
in Section \ref{s:relax}.
The combined computational secure protocol $AP_{2}$ is described in 
Section \ref{s:icsp}.
%The extended abstract is completed with conclusions
%and extensions. Proofs are only sketched in this extended abstract.
The improved resistant against Intruder-in the-Middle 
Attack $\widetilde{AP}_{2}$ protocol is introduced in Section \ref{s:IIMA}. 
The Conclusions in Section \ref{s:ConclusionsAndExtensions} complete the paper.

\section{Security Model for RFID Tags}
\label{s:fsd}

We consider the (RFID) Tag and the (RFID) Verifier.
 \remove{denoted by {\it S} and {\it R}, respectively.} 
The Tag and the Verifier communicate by sending and receiving messages according to their predefined programs, that form together a communication protocol.
We denote the $i^{th}$ message sent by the Tag and by the Verifier as $s_i$ and $r_i$, respectively. The sequence of alternating messages $M=s_1,r_1,s_2,r_2, \cdots$ sent during the course of the protocol execution can be divided into non overlapping subsequences, so that each subsequence $S_{i}={s_{i_k}, r_{i_k}}$ is called {\em communication session}. 
The union of the communication sessions forms the entire sequence of messages $M$.
Each $S_i$ starts with a message sent by the Tag and ends when the Verifier decides to send a message $r_{i_k}$=\textit{Open} or $r_{i_k}$=\textit{DoNotOpen}. In fact, the \textit{Open} message can be viewed as an electrical signal to the door of the car. Any message $s_{k}$ sent by the RFID Tag is defined as a key message.
Actually, the message $r_{i_k}$ represents a change in the state of the Verifier which corresponds to the Tag authentication as the one that may enter to use a resource. 

We assume a {\em Byzantine adversary} that listens in part or in all of sequence $M$ and may try to send complete messages on behalf of the Tag. The goal of the adversary is either making the Verifier send message $r=Open$ or driving the Verifier into a state after which the Verifier will not send the message $r=Open$ to the Tag. 
Given the features of the proposed model, we describe basic and combined authentication protocols. The first basic authentication protocol $AP_{1}$ is the proactive information theoretic secure protocol. The information theoretic security feature of this protocol is provided by the assumption that within any $n$ consecutive communication sessions $S_{i_{1}}=s_{i_1},r_{i_1}, \cdots S_{i_{n}}=s_{i_n}, r_{i_n}$ there is at least one message $s_{i_{k}}$ sent by the RFID Tag which the adversary is not aware of. The strict limitation imposed on the adversary is relaxed in the combined computational secure protocol $AP_{2}$. The security power of $AP_{1}$ and $AP_{2}$ protocols is based on random numbers generation and their updating at each communication session. 
$AP_{1}$ and $AP_{2}$ are introduced and analyzed in the next sections.  

\section{Proactive Information Theoretic Secure Protocol} \label {s:isp}

\par The proactive theoretic information secure protocol $AP_{1}$ is described in Figure \ref{f:SAP}.
Denote the accumulated random vector as $ARV[1.. n]$ and the last random vector that updates $ARV$ vector during the $i-th$ communication session as $LRV^{i}[1.. n]$. 

At the initialization stage the Tag and the Verifier both get a unique vector 
$ARV[1.. n]$ (lines 1-2).
In order to perform the authentication procedure, the Tag starts the communication session and passes to the Verifier the key message
$s_{1}=(ARV[n],LRV^{1}[1.. n]) $ (lines 6-8, Protocol for RFID Tag). 
%$s_{1}$ consists of the following pair: 
%vector's $ARV^{1}[1.. n]$ $n^{th}$ entry $ARV^{1}[n]$ and randomly generated $n$-dimensional vector 

After transmitting the first key message $s_{1}$,
the Tag and the Verifier, respectively, initialize $ARV[n]$ to zero and update $ARV[1.. n]$ vector by calculating $xor$ of each entry with the corresponding entry of $LRV^{1}[1.. n]$. 

%shift $B's$ rows below so that
%$b_{1}=(b_{1 1},b_{1 2},\ldots, b_{1 n})$ is treated as the first $B's$ row and the last row is deleted (lines 10-11 in Figure \ref{f:SAP}, Protocol for RFID Tag). 

During the next authentication session the Tag and the Verifier repeat the same
procedure: the Tag generates a new random $n$-dimensional vector\\ 
$LRV^{2}[1.. n]$,
and sends the newly generated key message \\
$s_{2}=(ARV[n-1],LRV^{2}[1.. n]) $, where the value in $ARV[n-1]$ serves as the key. 
 \remove{$s_{2}=(X_{2}, b_{2})$ to {\it R}.}
 \remove{ Here $X_{2}=a_{1 n-1}$, where 
$a_{1 n-1}= a_{1 n-1} \oplus b_{2 n-1}$.}
Then the new accumulated random vector $ARV[1..n]$ is equal to the $xor$ of the accumulated random vector used at the previous communication session, with the newly generated new random vector $LRV^{2}[1.. n]$.
% and the corresponding $ARV[1..n]'s$ entry sent by Tag to Verifier, is $ARV[n-1]$ entry.

%the $(n-1)-th$ entry of the accumulated random vector $ARV[1..n]$ sent in the second authentication message $s_{2}$ is equal to $(n-1)-th$ entry of the $xor$ ARV^{2}[n-1]$=AccumRV^{1}[n-1]\oplus LastRandom^{1}[n-1]$.
  
The Verifier generates the response message $r_{2}$ as either $Open$ or $DoNotOpen$ (lines 8, 10).
  
The authentication procedure is repeated continually scanning the vector $ARV[1..n]$ entries (one after the other) and updating $ARV[1..n]'s$ entries by initializing the lastly sent value to $0$ and calculating $xor$ of its entries with the corresponding entries of the newly randomly generated vector. After each $i^{th}$ authentication success both 
the Tag and the Verifier, respectively, initialize the $ARV[1.. n]$ entry used in $i-th$ authentication session, to $0$. The vector $ARV[1.. n]$ is updated by calculating $xor$ of each entry with the corresponding entry of the vector $LRV^{i}[1.. n]$. Note that $LRV^{i}[1.. n]$ has been previously randomly generated by the Tag and has been sent to the Verifier in the message $s_{i-1}$. The updating procedure and calculation of $xor$ for the corresponding $ARV[1.. n]$ entry are described in lines u1-u5.

In order to confirm the correct authentication, the RFID Verifier executes the authentication procedure in the following manner: upon receiving the key message $s_{i}=\left(ARV[n-(i-1)], LRV^{i}[1.. n]\right)$ the Verifier verifies that $ARV[n-(i-1)]'s$ value is the correct $(n-(i-1) (mod (n)))^{th}$ entry. If so, the Verifier confirms the correct authentication, ``transmits'' to the Tag the message $r_{i}=Open$ and updates the vector $ARV[1.. n]$ (lines 4-9, Protocol for RFID Verifier). Otherwise, the Verifier ``transmits'' to the Tag the message $r_{i}=DoNotOpen$ and does not update the vector $ARV[1.. n]$.

Assume that during the course of executing $AP_{1}$ it holds that in any sequence of alternating messages $M=s_1,r_1,s_2,r_2,\ldots $ the following condition is satisfied: in any $n$-length sequence \textit{M} of alternating messages between the Tag and the Verifier there is at least a single message $s_{{j_k}}$ not captured by the adversary.
Assume that in order to break the security system of the RFID Verifier, the adversary performs authentication procedure on behalf of the RFID Tag. To do so in any $S_{j}^{th}$ communication session the adversary has to forge the key message $s_{j_{i}}$, namely, to correctly guess the value of the corresponding $(n-(j_{i}-1) (mod (n)))^{th}$ entry of  $ARV[1.. n]$.

%Assume that $dim ~ARV[1.. n]=n$. 
Assume that the single unknown to the adversary key is the $n^{th}$ 
entry of $ARV[1.. n]$, namely $ARV[n]$ and the appropriate vector is
$LRV^{1}[1.. n]$ that has been sent by the Tag in the message $s_{1}=(ARV[n], LRV^{1}[1.. n])$ during the first communication session. 
%(Figure \ref{f:SAP2}, Step 1).

	After transmitting the first key message $s_{1}$
% $LastRandom^{1}[1.. n])$  to the RFID Verifier both \remove{\textit{S} and \textit{R}
the Tag and the Verifier update the vector $ARV[1.. n]$ according to the updating procedure described in lines u1-u5. 

\begin{figure*}
\begin{footnotesize}
\centering
% xxx
\begin{tabular}{|p{2.4in}||p{2.4in}|}
\hline&\\
\begin{minipage}[t]{2.1in}
\centering
{\it\bf Protocol for RFID Tag}
\begin{tabbing}
X: \= d \= d \= d \= d \= d \= d \= d \= d \= \kill 
1:\>\> \textbf{\textit{Initialization:}}\\
2:\>\> Create int array $ARV[1 .. n]$ \\  
3:\>\> int $i:=1$; \\
4: \>\> \textbf{Upon user request} \\
5:  \>\>   $keyentry=n-(i-1)mod~n$ \\
6: \>\> \>\> Create new random array $LRV$\\
%6: \>\> \>\> $X=a[keyentry]$\\
7: \>\> \>\> Send $s=(ARV[keyentry],LRV)$ \\
8:   \>\> \>\> \>\> to Verifier\\
9:\>\> \>\> If $Open$ received \\
10:\>\> \>\> \>\> Call \textbf{\textit{Updating procedure}}\\ 
11:\>\>  End user request \\
  \>\> \\ 
u1:\>\>  \textbf{\textit{Updating procedure}} \\ 
u2:\>\> \>\> $ARV[keyentry]=0$\\ 
u3:\>\> \>\> for all \ $ j ~ 1\leq j \leq n$ \\
u4:\>\> \>\> $ARV[j]=ARV[j] \oplus LRV[j]$ \\
u5:\>\> \>\> $i:=i+1$ \\
\>\> \\
 \end{tabbing}
\end{minipage}
&

\begin{minipage}[t]{2.1in}
\centering
{\it\bf Protocol for RFID Verifier}
\begin{tabbing}
X: \= d \= d \= d \= d \= d \= d \= d \= d \= \kill 
1:\>\> \textbf{\textit{Initialization:}}\\ 
2:\>\> Create int array $ARV[1 .. n]$ \\  
3:\>\> int $i:=1$; \\
4:\>\> \textbf{\textit{Upon reception of key message}} \\
5:\>\>   $keyentry=n-(i-1)mod~n$ \\
6:\>\> $s=(X,LRV)$ \\           
7:\>\> \>\> if $X=ARV[keyentry]$ \\ 
8:\>\> \>\> \>\> Send $Open$ and\\
9:\>\> \>\> \>\> Call \textbf{\textit{Updating procedure}} \\ 
10:\>\> \>\> else Send $DoNotOpen$\\
11:\>\> End of key message reception \\
  \>\> \\
u1:\>\>  \textbf{\textit{Updating procedure}} \\ 
u2:\>\> \>\> $ARV[keyentry]=0$\\ 
u3:\>\> \>\> for all $ j ~ 1\leq j \leq n$ \\
u4:\>\> \>\> $ARV[j]=ARV[j] \oplus LRV[j]$ \\
u5:\>\> \>\> $i:=i+1$ \\
\>\> \\
\end{tabbing}

\end{minipage}\\[1ex]
\hline
\end{tabular}
%0\hline&
\captionstyle{center}
\caption{Proactive Information Theoretic Secure Protocol $AP_{1}$.} 
\label{f:SAP} \end{footnotesize}
\end{figure*}
	
Note that in the next trial the Tag will send to the Verifier the updated \\
$(n-1)^{th}$ $ARV's$ entry that is equal to
	$ARV[n-1]$ used in the previous communication session, $xor-ed$ with $(n-1)-th$ entry of the new last
	randomly generated vector $LRV^2$.
	 
	Now vector $ARV[1.. n]$ is equal to the previous one
% $a_{1}$
with the initialized entry $ARV[n-1]=0$ $xor-ed$ with the vector $LRV^2[1.. n]$.
%$b_{2}=(b_{2 1},b_{2 2},\ldots, b_{2 n})$. 
The vector $ARV[1.. n]$ updating is done by the Tag and the Verifier
in each successful communication session. 

The $AP_{1}$ authentication protocol is information theoretic secure. This means that the probability that the adversary will forge the key message and perform the communication session on behalf of the RFID Tag successfully, is $2^{-l}$ which is negligible for a long enough $l$, where $l$ is the number of bits of the entry in the vector $ARV$.

The following Theorem proves that the introduced protocol is information theoretic secure.\\
\\ \noindent{\bf Theorem 1}  $AP_{1}$ protocol is information theoretic secure with a security parameter $l$ that is related to the number of bits in a key message, and proactive under the following assumptions: \\
$(i)$ The information exchanged during at least one of every $n$ successive communication sessions is not exposed to an adversary;\\
$(ii)$ Each session is atomic.

%\begin{sketch}
\noindent{\bf\noindent Proof:}
The $AP_{1}$ information security feature is based on the fact that at any authentication step $i$ the following conditions hold: (a) the RFID Tag and the RFID Verifier maintain the same vector $ARV[1.. n]$; (b) The Tag and the Verifier are synchronized in the sense that both the Tag and the Verifier perform the authentication procedure using as a key the same $n-(i-1) (mod (n))$ entry; (c) the vector $ARV[1.. n]$ shared by the Tag and the Verifier contains at least one entry unknown to the adversary.

 The proof continues by induction over the session number $i$.

\noindent{\bf Basis of induction $i=1$:}

(a) As it has been mentioned above, the first key message\\
 $s_{1}=(ARV[n],LRV^{1})$ at the first communication session $S_{1}$ contains $ARV[n]$ that is unknown to the adversary. Evidently, the Tag and the Verifier maintain the same vector $ARV[1.. n]$ that has been defined at the initialization stage when the adversary was not present.

(b) The Tag and the Verifier are synchronized because the first key message that the Tag sends to the Verifier and that the Verifier expects to receive is $ARV[n]$ which is the $n^{th}$ entry of the vector $ARV[1.. n]$ .

(c) Due to the initialization procedure, the vector $ARV[1.. n]$ granted to both the Tag and the Verifier, is entirely unknown to the adversary. 

\noindent{\bf Induction step:}
(a) Assume that during every $i<n$ communication sessions the Tag and the Verifier maintain the same vector $ARV[1.. n]$. Then the vector $ARV[1.. n]$ shared by the Tag and the Verifier during the next $i, i\geq n$ communication session will differ 
from the previous one by appropriate initializing of the used 
$n-(i-1) (mod (n))-th$ entry of $ARV[1.. n]$ 
and respective $xor-ing$ of each $ARV[1.. n]-th$ entry with the corresponding 
entry of the vector $LRV^{i}$ that has been sent to Verifier in the 
previous communication session. 

(b) Assume that during any $i<n$ communication session the Tag and the Verifier agree on the same 
$ARV[1.. n]^{th}$ entry $n-(i-1) (mod (n))$ that is the basis for constructing the key message. Then, at the next $(i+1)^{th}$ communication session the entry number is reduced by $1~ mod (n)$. As a result, the basis for constructing the key message at the Tag and the Verifier' sides, respectively, is the same $ARV's$ $n-(i-2) (mod (n))$ entry.

(c) For $i<n$ all the entries of the $ARV$ vector in each communication session $S_{j}$ among $i$ communication sessions $S_{1}, \ldots, S_{i}$ are unknown to the adversary. The induction assumption is correct due to the initialization procedure performed by the Tag and the Verifier, respectively. In addition, for any $i\geq n$ the basic condition that for each $i^{th}$ communication session $ARV[1.. n]^{th}$  entries are unknown to the adversary also holds. It is based on the assumption that among any $n$ successive communication sessions there is at least a single session that the adversary was not eavesdropping. 

Let us prove the information theoretic feature of $AP_{1}$. 
Assume, that among $n$ successive communication sessions $S_{i},S_{i+1},... S_{i+n-1}$ the adversary was aware of a certain $S_{i+k}$ session. In order to provide the authentication procedure on behalf of the Tag during any insecure communication session $S_{i+k+j}$ from $n-1$ following insecure sessions $S_{i+k+1}, S_{i+k+2},... ~S_{i+k+n-1}$ sessions, the adversary has to correctly guess the $n-(i+k-1)mod ~n$ entry of the vector $LRV^{i+k}[1..n]$ that securely refreshed $ARV[1.. n]$ of the Tag and the Verifier, respectively during $S_{i+k}$. Assuming the uniform distribution of the bits in the entry, the probability that the adversary will correctly guess the $n-(i+k-1)$ entry, is equal to $2^{-l}$, where $l$ is the number of bits in the entry.  
%In order to provide the authentication procedure on behalf of the Tag during any $S_{i}-th$ communication session, the RFID adversary has to correctly guess 

 The $AP_{1}$ proactive feature is proven in the following way. Assume that the adversary has gotten access to the whole vector $ARV[1.. n]$. Assume that in the $j^{th}$ communication session $S_{j}$ that follows this security failure, the adversary was not listening in to the message $s_{j}$ sent by the RFID Tag. In essence, during any of the following $(j+i)^{th}$ session, $i\geq 1$ each $ARV[1.. n]^{th}$ entry is $xor-ed$ with corresponding entry of the $LRV^{j}$. Note that the adversary was not listening in $LRV^{j}$.
 % contains at least a single number $b_{j+i, n-(j+i)+1}\in b_{j}$ that the adversary was not listening in. Here $b_{j}$ is the unknown to the adversary random vector that has been sent by \textit{S} in the secure message $s_{j}$. 
 Therefore, the basic condition, that within $n$ consecutive messages sent from the Tag to the Verifier there is at least a single message unknown to the adversary, is restored. As a result, the information theoretic security feature of $AP_{1}$ is regained.

Assume that the adversary tends to drive the RFID Verifier to a deadlock state after which the Tag will not be able to cause the Verifier to send a message \textit{r=Open}. Due to the session atomicity assumption, in order to do so the adversary must corrupt the vector $ARV$, say, by inserting a new value in $ARV$ entry on behalf of the RFID Tag. Nevertheless, the adversary will fail in this attempt because in order to insert a new entry in the vector $ARV$ the adversary has to authenticate himself or herself on behalf of the RFID Tag. The message $s_{j}$ that the adversary has to send to the Verifier must include the correct $ARV's$ entry.
%\end{sketch}

\qed

As a matter of fact, $AP_{1}$ has two parameters. The first parameter is vector' $ARV$ size $n$. The larger $n$ is, the weaker the assumption about the adversary is. The price paid for large $n$ is the additional memory used in the restricted memory size of the RFID devices. The second secure parameter is the number of bits $l$ of an entry in $ARV$. The longer $ARV's$ entries are, the smaller the probability for the adversary to guess the correct key is.

Note that when the assumption concerning one session in each sequential session, in which the adversary does not listen in, is violated, then the adversary can drive the system into a deadlock by, say, replacing $ARV's$ entries, by entries unknown to the Verifier. 

\section{Generalizing the Private Sessions Definition}
\label{s:relax}
This section generalizes the $1$ out of $n$ private communication
session assumption. Consider the cases in which $k \geq 1$ out of 
$n$ successive sessions are private, namely the adversary is not listening
in $k$ out of any $n$ successive sessions. 
In such cases the number of random numbers sent in
each communication session may be reduced. First we prove a lower bound
on the total number of random numbers which should be sent
during $n$ successive sessions. 

For proving a lower bound on the number of random numbers that should be sent during $n$ successive sessions, consider schemes for which
the vector-entries that are chosen to be {\it refreshed} by
random numbers which are specified by a deterministic 
function. A vector entry is {\em refreshed} by $xor-ing$ a 
new random number to the current vector entry 
or assigning the entry by a random number. 
We show that at least $n \cdot (k+1)$ new random numbers 
should be used during any $n$ successive communication sessions.

Consider any $n$ successive communication sessions. 
There are $n-k$ sessions in which the adversary may listen
in. Since we assume that the adversary knows the scheme,
the scheme must introduce at least $n-k+1$ refreshes for 
each vector-entry between any two successive usages of a vector-entry.
Thus the total number of refreshes in $n$ successive
sessions is at least $n \cdot (n-k+1)$ which implies at least $n-k+1$ or more refreshes in a single session.

The above lower bound is based on deterministic choices
of a refresh sequence which is known to the adversary.
In fact it is possible to use a {\em randomized scheme}, 
in which the vector-entries that are chosen to be {\it refreshed} by
random numbers, are {\it randomly} chosen.
Assume that the adversary does not know the 
identity of the randomly chosen vector-entries 
that are refreshed during the communication sessions 
the adversary is not listening in.
We show that it is possible to send only $(2n/k) (\log n)$ 
random numbers in each session. Thus, for a given 
(say, bounded by a constant) fraction of private communication $pcf=n/k$, 
the number of random numbers that need
to be sent in $n$ successive communication
sessions, is reduced from $n \cdot (n-n/pcf+1)$ to 
$2n \cdot pcf \cdot \log n$. 
Note that when $pcf$ is a constant these numbers
are $O(n^2)$ and $O(n \log n)$, respectively.

The {\em randomized scheme} chooses in each communication session
$2 \log n$ vector-entries and sends $2 \log n$ random numbers to
be $xor-ed$ with the corresponding vector-entries, sending the indices
of the chosen vector-entries as well.
We show that each entry is refreshed with high probability during the
$k$ private communication sessions that immediately precede it.

We now show that the probability that at least one refresh 
for each vector-entry takes place, is close to 1.
The probability that a certain entry is not refreshed is 
less than $(1-1/n)^{2n \log n}$
(the inequality is due to the fact that during one communication 
session no vector-entry is refreshed twice).
Given that $(1-1/n)^{2n \log n} \leq e^{-2 \log n}=1/n^2$,
it holds that the probability that all vector entries are refreshed
is greater than $1-\Sigma_{i=1}^n 1/n^2=1-1/n$.

\section{Combined Computational Secure Protocol} \label {s:icsp} 

We now allow the adversary to listen in any session between the RFID Tag and the RFID Verifier.
Our purpose is to enhance the basic proactive information theoretic secure protocol $AP_{1}$.

\begin{figure*}
\begin{footnotesize}
\centering
\begin{tabular}{|p{2.4in}||p{2.4in}|}
\hline&\\
\begin{minipage}[t]{2.1in}
\centering
{\it\bf Protocol for RFID Tag}
\begin{tabbing}
X: \= d \= d \= d \= d \= d \= d \= d \= d \= \kill 
1:\>\> \textbf{\textit{Initialization:}}\\
2:\>\> Create int vector array $ARV[1.. n]$ \\  
3:\>\> int $i:=1;~ seed:=0 $ \\
4:\>\> int vector arrays $keywords$ \\
\>\> \\
5: \>\> Upon user request \\
6: \>\> \>\> Create new random array $LRV$\\
7:\>\> \>\> $keyentry=n-(i-1)mod~n$ \\
8: \>\> \>\> $X[keyentry]=ARV[keyentry]$\\
9:\>\> \>\> Create pseudo-random sequence \\
10:\>\> \>\> $prs$ of length $m$ from\\ 
11:\>\> \>\> $seed=X[keyentry]\oplus seed$ \\
12:\>\> \>\> $Y=(LRV || keyword)\oplus prs$ \\
13: \>\> \>\> Send $Y$ to Verifier\\
14:\>\> \>\> If $Open$ received \\
15: \>\> \>\> Call \textbf{\textit{Updating procedure}} \\ 
16:\>\> End user request\\
\>\> \\ \\
u1:\>\>  \textbf{\textit{Updating procedure}} \\ 
u2:\>\> \>\> $ARV[keyentry]=0$\\ 
u3:\>\> \>\> for all $ j ~ 1\leq j\leq n$ \\
u4:\>\> \>\> $ARV[j]=ARV[j] \oplus LRV[j]$ \\
u5:\>\> \>\> $i:=i+1$ \\
%\>\> \\
\end{tabbing} 
\end{minipage} & 
\begin{minipage}[t]{2.1in} 
\centering {\it\bf Protocol for RFID Verifier} 
\begin{tabbing}
X: \= d \= d \= d \= d \= d \= d \= d \= d \= \kill 
1:\>\> \textbf{\textit{Initialization:}}\\
2:\>\> Create int vector array $ARV[1 .. n] $ \\  
3:\>\> int $i:=1;~seed:=0 $ \\
4:\>\> int vector arrays $keywords$ \\
  \\ 
5:\>\>Upon key message $Y$ reception \\
6:\>\> \>\> $keyentry=n-(i-1)mod~n$ \\ 
7:\>\> \>\> Create pseudo-random sequence \\
8:\>\> \>\> $prs$ of length $m$ from \\
9:\>\> \>\> $seed=Y[keyentry]\oplus seed$ \\
10:\>\> \>\>$Z=Y \oplus prs $ \\
11:\>\> \>\> if $Z[(n+1).. m]\in keywords$\\
12:\>\>\>\> \>\> Send $Open$ and \\
13:\>\>\>\> \>\> Call \textbf{\textit{Updating procedure}}\\ 
14:\>\> \>\> else \\ 
15:\>\>\>\> \>\> Send $DoNotOpen$ \\
16:\>\> End of key message reception \\
\>\> \\ \\
u1:\>\>  \textbf{\textit{Updating procedure}} \\ 
u2:\>\> \>\> $ARV[keyentry]=0$\\ 
u3:\>\> \>\> for all $ j ~ 1\leq j \leq n$ \\
u4:\>\> \>\> $ARV[j]=ARV[j] \oplus Z[j]$ \\
u5:\>\> \>\> $i:=i+1$ \\
%\>\> \\
\end{tabbing}
\end{minipage}\\[1ex]
\hline
\end{tabular}
\captionstyle{center}
\caption{Proactive Computational Secure Protocol $AP_{2}$.} \label{f:ComSec} \end{footnotesize} \end{figure*}

As in the $AP_{1}$ case, both the Tag and the Verifier get the initial $n$-dimensional vector $ARV$ in the initialization stage (Figure \ref{f:ComSec}, lines 1-4). 
In addition $k$ (a small number much less than $2^{k}$) bits different 
commands $keywords$ are granted to the RFID Tag and the RFID Verifier, 
respectively. These commands will be executed by the Verifier upon the 
Tag authentication. In the sequel, when no confusion is possible 
the $keyword$ used in this paper is $Open$; the $DoNotOpen$ keyword is used to 
refer to the situation in which the $keyword$ is not a valid command.

During the first authentication session the Tag executes the following encryption procedure:
New vector row $LRV[1.. n]$ is also created as in the proactive information theoretic secure protocol case.  
The $n^{th}$ $ARV's$ entry 
$ARV[n]$ is used as a seed for the generation of the pseudo-random sequence $prs$ of length $m=n\cdot l+k$, where $k$ is the $keyword$ length and $l$ is the length in bits of each $ARV$ vector entry. See \cite{MOV96}, Chapter 12 for possible choices of the generation mechanism of the pseudo-random numbers.  
 
The Tag creates a new vector row $Y$ that should be sent to the Verifier in the first authentication message. $Y$ is equal to \textit{xor} of the previously generated pseudo-random sequence $prs$ with vector $LRV^{1}$ concatenated with the $keyword$: $Y_{1}=prs  \oplus (LRV^{1}\| keyword)$ (Figure \ref{f:ComSec} lines 5-12). Eventually, the secure information encapsulation is provided. The first key message sent from the Tag to the Verifier during the first communication session is $s_{1}=Y$ (Figure \ref{f:ComSec}, Protocol for RFID Tag, line 13). It is assumed that both the Tag and the Verifier know the pseudo-random sequence procedure that produces $prs$. 

Upon receiving the message $s_{1}=Y$ the Verifier decrypts $s_{1}$ by calculating
$Y  \oplus prs $. If the decrypted suffix of the string is equal to the predefined string $keyword$, then the Verifier authenticates the Tag and returns the message $r_{1}=Open$ to the Tag. The updating of the vector $ARV$ is provided by the prefix of the decrypted string as in the basic information theoretic secure protocol. Otherwise, the message $r_{1}=DoNotOpen$ is sent to the Tag (lines 5-16, RFID  Verifier). The Updating procedure is described in lines u1-u4.

	 % (Figure \ref{f:ComSec} ,Protocol for RFID Tag, lines 7-18). 
During any $i^{th}$ authentication session $S_{i}, i=1, 2, \ldots $ the message $s_{i}$ sent by the Tag equals the $xor$ of the pseudo-random sequence $prs$ with the updated $i^{th}$ stage accumulated random vector $ARV$ concatenated with the $keyword$ string. 
Here $prs$ is the pseudo-random sequence generated by the $seed=X[keyentry] \oplus seed$, while the initial $seed$ value is initialized to zero and $keyentry=n-(i-1)mod~n$. $LRV$ is a newly generated random vector that updates the 
vector $ARV$.
It should be noted that the $keyword$ and the one way function that generates the pseudo-random numbers can be known to the adversary. The computational security of the designed $AP_{2}$ protocol is provided by means of the random seed generation in each session. Moreover, the recursive reuse of the seeds used in the previous communication sessions enhances the security of $AP_{2}$ where the adversary never listens in.

As a matter of fact, the seed $X[1]$ used in the first communication session $S_{1}$ is unknown to the adversary. The reason is that the adversary had not been present at the initialization stage. Therefore, the initial $ARV's$ entries are not available for the adversary. The seed updating is performed continuously in each communication session. Hence, the adversary does not get enough time to guess the secret seeds by observing the transmitted messages.
  
In essence, the encryption scheme is based on the message encapsulation by means of the One Time Pads techniques (e. g., \cite {S02}), whereas the pads are created by pseudo-random sequence using a randomly created seed defined by the updating procedure of the vector $ARV$. The following theorem proves the correctness of $AP_{2}$. \\

\noindent{\bf Theorem 2}
The $AP_{2}$ protocol is proactive computationally secure under the assumption that each communication session is atomic. The security parameters are yield by the one way function used to produce the pseudo random sequence and by the the seed length.
%the adversary may listen in any communication session between Tag and Verifier.

\noindent{\bf\noindent Proof:}
Assume that the adversary is listening in all communication sessions $S_{i_{1}}, \ldots, S_{i_{n}}$ between the Tag and the Verifier. Even though the one way function $f$ which generates the pseudo-random sequence is available to the adversary, calculating its invert $f^{-1}$ is computationally infeasible \remove{(\cite {S02})}. Hence, correct prediction of the seed $X[i_{n+1}]$ and the corresponding pseudo-random sequence $c$ for the next communication session $S_{i_{n+1}}$ that the adversary wishes to provide in order to break the security system, is computationally infeasible. The probability of the adversarial success is determined by the probability to invert $f$ function.

The Verifier confirms the Tag authentication at each $i^{th}$ communication session by revealing the $keyword$ string from the received decrypted message $s_{i}$. If the decrypted $keyword$ string is correct, then the Verifier accepts the Tag's correct authentication. 

We now prove the proactive feature of $AP_{2}$. Assume that the adversary has successfully broken the security system and has gotten access to the whole vector $ARV$. Hence, the adversary can correctly calculate the seeds that should be used in the following sessions.
However, after the first session in which the adversary is not present, $AP_{2}$ satisfies the conditions of the information theoretic secure protocol $AP_{1}$. As a result, the information theoretic and computational security features are restored.

\qed

The $AP_{2}$'s parameters that define the pseudo-random sequence length are the number of entries of $ARV$- $n$, the number of bits of an entry in $ARV$- $l$, and the $keyword$ length- $k$. 

Note that it is possible to use the pseudo-random sequence only once
in every $n$ successive sessions, reducing the processing required
in the rest $n-1$ out of $n$ sessions. One may view the
session that uses the pseudo-random sequence as a computational
way to ensure that the adversary does not listen in (or able to
decrypt the communication) at least in one session in
every $n$ successive sessions. 

\section{Resistance Against Intruder-in the Middle-Attack}
\label{s:IIMA}

In this section we upgrade the computationally secure protocol $AP_{2}$ in order to be able to cope with the Intruder in the Middle Attack (IIMA), see e. g., \cite{S02}. This type of attack is possible when the intruder captures the encrypted messages sent by the RFID Tag to the RFID Verifier and
 \remove{changes the bits of the message even without trying to provide the authentication procedure on behalf of the Tag.}
uses the captured messages by replacing a modified version of them. Such IIMA may drive the protocol to a deadlock state.
We relax the assumption concerning the atomicity of each communication session coping with adversarial success in performing IIMA that may immediately lead the RFID Tag to change the basic vector $ARV$. As a result, the Verifier enters a deadlock state after which it will be unable to send the message $Open$. In order to strengthen the $AP_{2}$ protocol against the IIMA we propose to use digital watermarking \cite{DWM} and redundant coding \cite{rc}. Note that we do not use any cryptographic hash function as in the Message Authentication Code (MAC) schemes such as \cite{S02}. The extended computationally resistant against IIMA $\widetilde{AP}_{2}$ protocol is defined in the following way. As in the $AP_{2}$ case the encryption key is derived from the basic vector $ARV$. The seed $X_{j}$ calculated from the corresponding vector-entry and from the seeds used in the previous sessions, is divided now into four independent seeds $X_{j}^{1}, ~X_{j}^{2},~ X_{j}^{3}$, and $X_{j}^{4}$. Each seed $X_{j}^{k} , ~ k=1, \ldots, 4$, generates a corresponding pseudo-random sequence $c^{j_{k}}$. The RFID Tag implements the following encryption scheme:

Let $m$ be a total length in bits of the encapsulated encrypted message $s_{j}$ of $AP_{2}$, where $m=nl+k$ as was defined in Section \ref{s:icsp}, $v$ be the total number of the watermarks 
$w_{1},\ldots, w_{v} $ added to $s_{j}$, $d_{min}$ be a Hamming distance of the appropriate error detection code, and $q$ be the number of redundant bits $r_{1},\ldots, r_{q} $ used to extend the bits of the message defined by $AP_{2}$ to form a legal codeword. Actually, the total length of the key message $Y_{j}$ sent during any $j^{th}$ communication session is equal to $t=m+q+v$. 
The resulting $t$ bits message is sent during the $j^{th}$ authentication session. $s_{j}$ has the following structure: \\
 $Y_{j}= \pi_{X_{j}^{4}} ((LRV^{j}\|keyword)\oplus (c^{j_{1}}_{1},\ldots,c^{j_{1}}_{m})  \| (r_{1}, \ldots r_{q}) \oplus (c^{j_{2}}_{1},\ldots,c^{j_{2}}_{q})  \| (w_{1},\ldots, w_{v}) )$. Here 
$\pi_{X_{j}^{4}}$ determines the pseudo-random permutation of the concatenated string \\
 $((LRV^{j}\|keyword)\oplus (c^{j_{1}}_{1},\ldots,c^{j_{1}}_{m})  \| (r_{1}, \ldots r_{q}) \oplus (c^{j_{2}}_{1},\ldots,c^{j_{2}}_{q})  \| (w_{1},\ldots, w_{v}) )$.

The pseudo-random sequence $c^{j_{1}}$ encapsulates the newly generated random string $LRV^{j}$ concatenated with the $keyword$ string as in the $AP_{2}$ case. 
The basic random string $LRV^{j}$ concatenated with the $keyword$ string is extended by error detection redundancy bits to form a legal codeword. The redundant bits $r_{1}, \cdots, r_{q}$ are located after the sub-string $(LRV^{j}\|keyword)$ in the message.
% The pseudo-random sequence $c^{j_{1}}$ encapsulates the newly generated random string $LRV^{j}$ concatenated with the $keyword$ string as in the $AP_{2}$ case. 
The pseudo-random sequence $c^{j_{2}}$ generated from the seed $X_{j}^{2}$ encapsulates the redundant bits $r_{1}, \cdots, r_{q}$.
  The pseudo-random sequence $c^{j_{3}}$ generated from the seed $X_{j}^{3}$ determines the watermarks $w_{1},\ldots, w_{v}$ values that are located after the code redundant bits in the composed string message. $c^{j_{3}}$ is created as $v$ bits length sequence, while each watermark is 1 bit in length.
Finally, the pseudo-random sequence $c^{j_{4}}$ generated from the seed $X_{j}^{4}$ determines the pseudo-random permutation $\pi$  of the composed string that includes the string $(LRV^{j}\|keyword)$ encapsulated by $c^{j_{1}}$, redundant bits $r_{1}, \ldots, r_{q}$ encapsulated by $c^{j_{3}}$, and the unprotected watermarks. It should be remembered that $c^{j_{4}}$ should produce a permutation for $t=m+q+v$ bits length sequence (in fact $X^4_j$ may deterministically define a permutation as suggested in \cite{DLH07}).
% $c_{j_{4}}$ has the length of $L\cdot log ~L$ bits because $c_{j_{4}}$ has $L$ entries, while for representing a new index for a given entry $log~L$ bits should be used. 

%As the result, the key message $Y_{j}$ sent by $S$ to $R$ during any communication session $S_{j}$ is 
%$Y_{j}= \pi (b_{j}\|keyword[k] \| (r_{1}, \ldots r_{q}) \| (w_{1},\ldots, w_{v}) )$.
%Actually, from the adversarial point of view the strings concatenation \\
% $(b_{j} \| keyword[k])$ is mixed by the uniformly distributed watermarks and code's redundant bits. Note that the watermarks are sent in unprotected manner.

The advantage of this approach is that the original string $(LRV^{j} \| keyword)$ and the corresponding redundant bits $r_{1}, \cdots, r_{q}$ are encapsulated and, therefore protected in an independent way. The redundant code that can be effective in the key string protection against IIMA must have a sufficiently large Hamming distance \cite{rc}. Assume that the adversarial goal is to corrupt the key message and to change the transmitted vector that should update the vector $ARV$. In order to succeed in his/her attempt, the adversary must change the original string, that is, change a correct codeword, to another correct codeword string.  
% \cite{rc}.
The larger the code Hamming distance is, the smaller the probability for the adversary to succeed without changing watermarks.  

Any linear block code with a large Hamming distance may fit. The great advantage of linear codes is that they can be easily implemented in hardware based on Linear Feed-Back Registers \cite{rc}. Since our schemes are based only on $xor$ and pseudo-random sequences, we consider the code which is based on the composition of $log(nl)$ $xor$ checks \cite{CT}. This code is defined as the composition of $log (nl)$ parity checks while the redundant bits in each dimension are equal to the $xor$ of the corresponding bits of the $(LRV^{j} \| keyword)$ string. The Hamming distance of this composed code is equal to $log (nl)+1$ \cite{CT}. The code's construction is as follows: the original string is represented as the $log (nl)$-dimensional hypercube while the redundant parity check bits are added in each dimension. The overhead of the redundant bits is equal to 
$q=log(nl)\cdot \sqrt[log (nl)+1]{(nl)}^{log (nl)}$.      

The resistance against IIMA of the extended $\widetilde{AP}_{2}$ protocol is based on the following observations.

%\noindent{\bf Lemma 1}
%$\widetilde{AP}_{2}$ protocol is computationally secure against IIMA.\\
%\noindent{\bf\noindent Proof:}
Assume that the adversary has changed the bits of a certain message $s_{j}=Y_{j}$ that has been sent by the RFID Tag during the communication session $S_{j}$. Let us evaluate the probability $P_{A}$ of the adversarial success.

%Without loss of generality, assume that each entry of the key string $s_{j}$ is $l$ bits long.
%The following notations are introduced:

Assume that the encryption scheme is well known to the adversary. The unique information that is not recovered by the adversaey is the $X_{j}$ number and the seeds $X_{j}^{1}, ~X_{j}^{2},~X_{j}^{3}, ~X_{j}^{4}$ generated from it.

The seed $X_{j}^{4}$ produces a pseudo-random sequence; hence, from the adversarial point of view any bit has the same probability of being a watermark.
Therefore, the probability that the adversary will corrupt a watermark while changing the bits of $s_{j}$ is equal to $\alpha=\frac{v}{t}$. In order to successfully change the part of the original message $s_{j}$, the adversary has to corrupt at least $d_{min}$ bits of $s_{j}$ that are the random bits of $(LRV^{j} \| keyword)$. Based on the assumption concerning the uniform distribution of the watermark bits, the probability of the adversarial succeess is bounded by $P_{A}\leq \left(1-\alpha\right)^{d_{min}}$. $P_{A}$ may be as small as possible by choosing large enough vector $ARV$ dimension $n$, number of the artificially inserted watermarks $v$, and number of redundant bits $q$ used to obtain a large Hamming distance $d_{min}$ between any two codewords.  

%\qed

Note that there is a trade-off between the $n,~k,~v,~l,$ and $d_{min}$ values and minimization of $P_{A}$. Let us consider the following example. Assume that the artificially inserted watermarks occupy half of the encrypted message providing $\alpha=\frac{1}{2}$. Assume that the redundant code is the composition of $log(nl)$ $xor$-based parity check codes. Then the code minimal distance is $d_{min}=log (nl)+1$. The probability $P_{A}$ of the adversarial success is evaluated as:\\ $P_{A}\leq (\frac{1}{2})^ {log (nl)+1}=\frac{1}{2\cdot n \cdot l}$. For large enough $n$ and $l$, $P_{A}$ will be negligible.

\section{Conclusions and Extensions}
\label{s:ConclusionsAndExtensions}

%\section{Conclusions and Extensions}

We presented a secure authentication protocol that is based on the assumption that among any $n$ consecutive interactions between the RFID Tag and the RFID Verifier there is at least a single session in which the adversary was not listening in. This model is not perfect; nevertheless it takes into account the restricted capabilities of the real world RFID adversary. Actually, $AP_{1}$ provides information theoretic security guarantees.

The $AP_{2}$ protocol loosens the assumption of the RFID adversary's weakness. It provides computational security in a proactive manner. The computational security of $AP_{2}$ is provided by involving basic arithmetic operations and using small size memory. 
The larger are the vector' $ARV$ entries,
the generated pseudo-random sequence is closer to a real random sequence (\cite {MOV96}).

The updated protocol $\widetilde{ AP_{2}}$ provides computationally secure resistance also against IIMAs, loosening the session atomicity assumption. Its computational security power strictly depends on $n$ the size of the vector $ARV$, the overhead of the artificially inserted watermarks, and error detection power of the redundant code.

Note that one can use symmetric 
authentication scheme to obtain mutual authentication of the
Tag and the Verifier. For example, we may
double the number of entries in the vectors of the Tag
and the Verifier and use one entry to authenticate the
Tag and the next entry to authenticate the Verifier. The mutual authentication version may support production of random numbers by one or both sides. In case of one sided production (say by the Verifier) the random number that should be used by the Tag is sent to the Tag as part of the message from the Verifier.

If a specific application requires the Tag and the Verifier's synchronization, our protocols should be extended by the Automatic Repeat Request (ARQ) mechanism. The current session number should be sent by the Tag as part of the encrypted authentication message, and the Verifier has to ACK the reception of the authentication message with a certain number in an encrypted manner, as well. In order to keep the Tag and the Verifier in synchronous state, the session sequential number and the ARQ mechanisms should be carefully incorporated in the presented protocols.

The $AP_{1}$ and $AP_{2}$ protocols can be used in the case of multiple RFID Tags and a single RFID Verifier. In order to provide secure communication the RFID Verifier has to store different vectors and to share a unique vector with each RFID Tag. As a matter of fact, the limitations imposed on the number of RFID Tags are only related to the limited storage capabilities of the RFID Verifier.
%More details concerning the extensions above are deferred to the full version of the paper.

%It should be mentioned that it is possible to reduce the memory usage of the 
%RFIDs be a factor of two; the Tag and the receiver can perform the 
%%random values of the basic vector $ARV$. 
%Thus, only the next to use column of the vector includes entries for all
%rows, while, say, the most recently used column consists of only one row,
%with the single random number assigned in the very last session.

\noindent{\bf Acknowledgment.} We thank Ari Juels for helpful remarks.

 \end{document}